\documentstyle[12pt]{article}
\newcommand{\be}[1]{\begin{equation}\label{#1}}                     
\newcommand{\ba}[1]{\begin{eqnarray}\label{#1}}                     
\newcommand{\ee}{\end{equation}}                                    
\newcommand{\ea}{\end{eqnarray}}                                    
\newcommand{\non}{\nonumber\\\rule{0pt}{30pt}}

\newcommand{\dis}{\displaystyle}                                    
\newcommand{\Eq}[1]{(\ref{#1})}                                    
\makeatletter
\@addtoreset{equation}{section}
\makeatother

\begin{document}
%



\begin{center}
\begin{Large}
{\bf 
The New Identity for the Scattering Matrix of\\
Exactly Solvable Models%
\footnote{Dedicated to J.~Zittartz on the occasion of his 60th birthday}
}
\end{Large}

\vspace{36pt}

Vladimir Korepin\raisebox{2mm}{{\scriptsize $\dagger$}}
                    and 
Nikita Slavnov\raisebox{2mm}{{\scriptsize $\ddagger$}}

\vspace{20pt}

~\raisebox{2mm}{{\scriptsize $\dagger$}}
{\it ITP, SUNY at Stony Brook, NY 11794-3840, USA.}\\
korepin@insti.physics.sunysb.edu

\vspace{10pt}

~\raisebox{2mm}{{\scriptsize $\ddagger$}}
{\it Steklov Mathematical Institute,
Gubkina 8, Moscow 117966, Russia.}\\
nslavnov@mi.ras.ru\\

\vspace{42pt}
Abstract
\end{center}

{\small
We discovered a simple quadratic equation, which relates scattering phases
of particles on Fermi surface. We consider one dimensional Bose gas and
$XXZ$ Heisenberg spin chain.}
\vspace{10pt}

\newpage

\section{Introduction}

To define everything precisely we  consider a specific model.
Let us concentrate our attention on Bose gas with delta
interaction (quantum Nonlinear Schr\"odinger equation).

The Hamiltonian of the model is
\be{Hamilton}
{H}={\dis\int \,dx
\left({\partial_x}\Psi^{\dagger}(x)
{\partial_x} \Psi(x)+
c\Psi^{\dagger}(x)\Psi^{\dagger}(x)\Psi(x)\Psi(x)
-h \Psi^{\dagger}(x)\Psi(x)\right).}
\ee
Here $0<c<\infty$, $h>0 $ are the coupling constant and 
 the chemical potential respectively. The canonical Bose fields
$\Psi(x,t),~\Psi^{\dagger}(x,t),~(x,t \in {\bf R})$
obey  the standard commutation relations
\be{Icom}
[\Psi(x,t), \Psi^{\dagger}(y,t)]=\delta(x-y).
\ee
and act in the Fock space
with the vacuum vector $|0\rangle$, which
is characterized by the relation:
\be{Ivac}
\Psi(x,t)|0\rangle =0.
\ee

Alternatively the model can be formulated on the
language of many-body quantum mechanics system, containing
 $N$ identical particles. In this case the Hamiltonian of the Bose gas can be
 represented as
\begin{equation}\label{Ham1}
{\cal H}_N=-\sum_{j=1}^{N}\frac{\partial^2}{\partial x_j^2}
+2c\sum_{N\ge j>k\ge1}\delta(x_j-x_k)-hN.
\end{equation}
For nonzero value of the coupling constant the Pauli principal
is valid (chapter VII of \cite{KBI}). 

The model was solved by Bethe Ansatz \cite{LL}. The ground state is
a Fermi sphere. In order to describe it precisely it is convenient to
introduce spectral parameter $\lambda$ (similar to rapidity). The derivative
of the momentum of the particle with respect to the spectral parameter is
\begin{equation}\label{denom}
\frac{\partial k(\lambda)}{\partial\lambda}=2\pi \rho(\lambda),
\end{equation}
where the function $\rho(\lambda)$ is defined by an integral equation
\begin{equation}\label{rho0} 
\rho(\lambda)-\frac{1}{2\pi}\int_{-q}^{q} K(\lambda,\mu)
\rho(\mu)\,d\mu=\frac{1}{2\pi}.
\end{equation}
Here $q$ is the value of the spectral parameter on the Fermi surface, and
\begin{equation}\label{kernel}
K(\lambda,\mu)=\frac{2c}{c^2+(\lambda-\mu)^2}.
\end{equation}
One can prove that the integral operator $\hat I-\frac1{2\pi}
\hat K$ is not degenerated, and hence, the
equation \Eq{rho0} has unique solution (\cite{YY}, chapter I of \cite{KBI}).
 The density of the gas is given by
\be{totden}
D=\int_{-q}^q\rho(\lambda)\,d\lambda.
\ee
There is one particle in the model.
It  is defined at $\lambda \ge q$ or $\lambda\le -q$.
 The energy of the particle $\varepsilon(\lambda)$ is
\begin{equation}\label{eps0}
\varepsilon(\lambda)-\frac{1}{2\pi}\int_{-q}^{q} K(\lambda,\mu)
\varepsilon(\mu)\,d\mu=\lambda^2-h.
\end{equation}
It vanishes on the Fermi surface $\varepsilon(\pm q)=0$. 
The momentum is
\begin{equation}\label{mom0}
k(\lambda)=\lambda+\int_{-q}^{q} \theta(\lambda-\mu)
\rho(\mu)\,d\mu.
\end{equation}
Here
\begin{equation}\label{theta}
\theta(\lambda)=i\ln\left(\frac{ic+\lambda}{ic-\lambda}\right).
\end{equation}

One can calculate a scattering matrix of particle with spectral parameter
$\lambda$ on another particle with spectral parameter $\mu$ 
(chapter I of \cite{KBI}). There
is no multi-particle production or reflection. Transition coefficient
is
\be{trancoef}
\exp\{2\pi iF(\lambda|\mu)\}.
\ee
The phase $F(\lambda|\mu)$ is defined by an integral equation
\begin{equation}\label{phase}
F(\lambda|\mu)-\frac{1}{2\pi}\int_{-q}^{q} K(\lambda,\nu)
F(\nu|\mu)\,d\nu=\frac 1{2\pi}\theta(\lambda-\mu).
\end{equation}

The most important are scattering phases of particles on the Fermi
edges $F(q|q)$ and $F(q|-q)$. In this paper we prove the identity
\be{ident2}
\det\left(
\begin{array}{cc}
1-F(q|q)&F(q|-q)\\
-F(-q|q)&1+F(-q|-q)
\end{array}\right)
=1.
\ee
This is the main result of the paper. Another way to rewrite this
identity is
\be{ident1}
\Bigl(1-F(q|q)\Bigr)^2-F^2(q|-q)=1.
\ee
Here we have used the property $F(-\lambda|-\mu)=-F(\lambda|\mu)$, which
follows immediately from the antisymmetry of $\theta(\lambda-\mu)=
-\theta(\mu-\lambda)$.

This identity also permit us to relate ``fractional'' charge to the phase
shift on the Fermi surface. Fractional charge ${\cal Z}$ appears in formul\ae~
for finite size corrections
(chapter I of \cite{KBI}). This value is necessary for conformal description
of the model (chapter XVIII of \cite{KBI}) and it is equal to
\be{confch}
{\cal Z}=2\pi \rho(q).
\ee
Using the equations \Eq{rho0}, \Eq{phase} for $\rho(\lambda)$ and
$F(\lambda|\mu)$, one can find the relationship between the
fractional charge and the scattering phase on the Fermi surface
\be{chph1}
{\cal Z}=1+F(q|-q)-F(q|q).
\ee
Indeed, it follows from \Eq{rho0} that
\begin{equation}\label{rho1} 
[2\pi\rho(\lambda)-1]-\frac{1}{2\pi}\int_{-q}^{q} K(\lambda,\mu)
[2\pi\rho(\mu)-1]\,d\mu=\frac{1}{2\pi}[\theta(\lambda+q)-
\theta(\lambda-q)].
\end{equation}
Comparing this equation with \Eq{phase} we find
\be{auxr}
2\pi\rho(\lambda)=1+F(\lambda|-q)-F(\lambda|q),
\ee
what, in turns, implies \Eq{chph1}.
The identity \Eq{ident1} allows us to find new relation
\be{chph2}
{\cal Z}^{-1}=1-F(q|-q)-F(q|q).
\ee
%

\section{The proof of the main identity}

In this section we give the proof of the identity \Eq{ident2}.
In order to do this, one should calculate the derivatives of the
function $F(\lambda|\mu)$ with respect to $\lambda$, $\mu$ and $q$.
Using the basic equation \Eq{phase}, we have
\ba{derivl}
&&{\dis\hspace{-1cm}
\frac{\partial F(\lambda|\mu)}{\partial\lambda}
-\frac{1}{2\pi}\int_{-q}^{q} K(\lambda,\nu)
\frac{\partial F(\nu|\mu)}{\partial\nu}\,d\nu=
\frac 1{2\pi}K(\lambda,\mu)}\non
&&{\dis\hspace{2.5cm}-
\frac 1{2\pi}K(\lambda,q)F(q|\mu)+
\frac 1{2\pi}K(\lambda,-q)F(-q|\mu),}
\ea
\vskip5mm
\be{derivm}
\hspace{-1cm}\frac{\partial F(\lambda|\mu)}{\partial\mu}
-\frac{1}{2\pi}\int_{-q}^{q} K(\lambda,\nu)
\frac{\partial F(\nu|\mu)}{\partial\mu}\,d\nu=
-\frac 1{2\pi}K(\lambda,\mu),
\ee
\vskip5mm
\ba{derivq}
&&{\dis\hspace{-1cm}
\frac{\partial F(\lambda|\mu)}{\partial q}
-\frac{1}{2\pi}\int_{-q}^{q} K(\lambda,\nu)
\frac{\partial F(\nu|\mu)}{\partial q}\,d\nu}\non
&&{\dis\hspace{2.5cm}=
\frac 1{2\pi}K(\lambda,q)F(q|\mu)+
\frac 1{2\pi}K(\lambda,-q)F(-q|\mu).}
\ea
Here we have used that $\frac{\partial}{\partial\lambda}
\theta(\lambda-\mu)=K(\lambda,\mu)$.

As we have mentioned already,
the resolvent of the operator $\hat I-\frac1{2\pi}\hat K$
exists and it is equal to
\be{resolv}
\hat R=\left(\hat I-\frac1{2\pi}
\hat K\right)^{-1}\cdot \frac1{2\pi}\hat K.
\ee
The derivatives of the function $F(\lambda|\mu)$ can be expressed in terms
of the resolvent
\be{deriv}
\begin{array}{l}
{\dis
\frac{\partial F(\lambda|\mu)}{\partial\lambda}
=R(\lambda,\mu)-
R(\lambda,q)F(q|\mu)+R(\lambda,-q)F(-q|\mu)}\non
{\dis
\frac{\partial F(\lambda|\mu)}{\partial\mu}
=-R(\lambda,\mu)}\non
{\dis
\frac{\partial F(\lambda|\mu)}{\partial q}
=R(\lambda,q)F(q|\mu)+R(\lambda,-q)F(-q|\mu)}
\end{array}
\ee
Using these equations one can find the complete derivatives with respect
to $q$ of functions $F(q|q)$,  $F(q|-q)$ etc.:
\be{comder0}
\begin{array}{l}
{\dis
\frac d{dq}F(q|\pm q)=\left.\left(
\frac\partial{\partial\lambda}\pm
\frac\partial{\partial\mu}+\frac\partial{\partial q}\right)F(\lambda|\mu)
\right|_{\lambda=q\atop{\mu=\pm q}},}\non
{\dis
\frac d{dq}F(-q|\pm q)=\left.\left(
-\frac\partial{\partial\lambda}\pm
\frac\partial{\partial\mu}+\frac\partial{\partial q}\right)F(\lambda|\mu)
\right|_{\lambda=-q\atop{\mu=\pm q}}.}
\end{array}
\ee
Substituting here equations \Eq{deriv} we find
\be{compder}
\begin{array}{l}
{\dis
\frac d{dq}F( q|- q)=
2R(q,-q)\Bigl(1+
F(- q|-q)\Bigr),}\non
{\dis
\frac d{dq}F(- q| q)=
- 2R(- q, q)\Bigl(1-
F(q| q)\Bigr),}\non
{\dis
\frac d{dq}F( q| q)=
 2R( q,- q)F(- q|q),}\non
{\dis
\frac d{dq}F(- q|-q)=
 2R(-q, q)F( q|-q).}
\end{array}
\ee

Now it is sufficient to take the derivative with respect to $q$ of the
l.h.s. of the equation \Eq{ident2}: 
\ba{derident}
&&{\dis\hspace{-5mm}
\frac d{dq}\det\left(
\begin{array}{cc}
1-F(q|q)&F(q|-q)\\
-F(-q|q)&1+F(-q|-q)
\end{array}\right)}\non
&&{\dis\hspace{5mm}=
-\frac{dF(q|q)}{dq}\Bigl(1+F(-q|-q)\Bigr)+
\Bigl(1-F(q|q)\Bigr)\frac{dF(-q|-q)}{dq}}\non
&&{\dis\hspace{1cm}+
\frac{dF(q|-q)}{dq}F(-q|q)+F(q|-q)\frac{dF(-q|q)}{dq}}\non
&&{\dis\hspace{5mm}=-
2R(q,-q)F(-q|q)\Bigl(1+F(-q|-q)\Bigr)}\non
&&{\dis\hspace{13mm}+
2R(-q,q)F(q|-q)\Bigl(1-F(q|q)\Bigr)
}\non
&&{\dis\hspace{21mm}+
2R(q,-q)F(-q|q)\Bigl(1+F(-q|-q)\Bigr)}\non
&&{\dis\hspace{29mm}-
2R(-q,q)F(q|-q)\Bigl(1-F(q|q)\Bigr)
=0.}
\ea
On the other hand, it is clear that for $q=0$ we have
\be{ident0}
\left.\det\left(
\begin{array}{cc}
1-F(q|q)&F(q|-q)\\
-F(-q|q)&1+F(-q|-q)
\end{array}\right)
\right|_{q=0}
=1.
\ee
Thus, the identity \Eq{ident2} is proved.

We would like to emphasize the we did not use the explicit
expressions for the kernel $K(\lambda,\mu)$ and the function
$\theta(\lambda-\mu)$. In fact, we have used only three properties:

a) the existence of the resolvent of the operator 
$\hat I-\frac1{2\pi}\hat K$;

b) the kernel $K(\lambda,\mu)$ and the function
$\theta(\lambda-\mu)$ depend on the difference;

c)  the derivative of $\theta(\lambda-\mu)$ is equal to the
kernel $K(\lambda,\mu)$.
\vskip5mm

In order to reduce \Eq{ident2} to the identity \Eq{ident1} one
should use also the antisymmetry property $\theta(\lambda-\mu)=
-\theta(\mu-\lambda)$.

Thus, the quadratic identity for the scattering phase is valid for a
wide class of completely integrable models, but not only for 
the one-dimensional Bose gas. In particular, it is valid for scattering
phases of elementary particles (spin waves) of $XXZ$ Heisenberg spin
chain in a magnetic field (chapter II of \cite{KBI}).

\section*{Acknowledgments}
This work was  supported by the National Science Foundation (NSF)
Grant No. PHY-9321165, and the Russian Foundation of Basic Research
Grant No. 96-01-00344.


\begin{thebibliography}{99}
%
\bibitem{KBI}V.~E.~Korepin, N.~M.~Bogoliubov and
A.~G.~Izergin,
Quantum Inverse Scattering Method and Correlation
Functions (Cambridge University Press, 1993)
%
\bibitem{LL}E.~H.~Lieb and  W.~Liniger,
 Phys. Rev. {\bf 130},1605 (1963)
%
\bibitem{YY}C.~N.~Yang and C.~P.~Yang,
J. Math. Phys. {\bf 10},1115 (1969)
%
\end{thebibliography}
\end{document}